\documentclass[signlecolumn,preprint]{aastex63}

\usepackage{url}

\usepackage{xcolor} 

\received{xxx xx, 2020}
\revised{xxx xx, 2020}
\accepted{xxx xx, 2020}

\submitjournal{ApJ}

\shorttitle{Small Sensitivity to Model Resolution}
\shortauthors{Wei, Zhang, and Yang}

\graphicspath{{./}{figures/}}

\begin{document}

\title{Small Sensitivity of the Simulated Climate of Tidally Locked Aquaplanets to Model Resolution}

\correspondingauthor{Jun Yang}
\email{junyang@pku.edu.cn}

\author[0000-0002-0588-6600]{Mengyu Wei}
\affiliation{Department of Atmospheric and Oceanic Sciences, School of Physics, Peking University, Beijing 100871, China.}
\affiliation{School of Oceanic and Atmospheric Sciences, The Ocean University of China, Qingdao 266100, China.}

\author[0000-0001-6194-760X]{Yixiao Zhang}
\affiliation{Department of Atmospheric and Oceanic Sciences, School of Physics, Peking University, Beijing 100871, China.}

\author[0000-0001-6031-2485]{Jun Yang}
\affiliation{Department of Atmospheric and Oceanic Sciences, School of Physics, Peking University, Beijing 100871, China.}

\begin{abstract}

Tidally locked terrestrial planets around low-mass stars are the prime targets of finding potentially habitable exoplanets. Several atmospheric general circulation models have been employed to simulate their possible climates, however, model intercomparisons showed that there are large differences in the results of the models even when they are forced with the same boundary conditions. In this paper, we examine whether model resolution contributes to the differences. Using the atmospheric general circulation model ExoCAM coupled to a 50-m slab ocean, we examine three different horizontal resolutions (440~km\,$\times$\,550~km, 210~km\,$\times$\,280~km, and 50~km\,$\times$\,70~km in latitude and longitude) and three different vertical resolutions (26, 51, and 74 levels) under the same dynamical core and the same schemes of radiation, convection and clouds. Among the experiments, the differences are within 5 K in global-mean surface temperature and within 0.007 in planetary albedo. These differences are from cloud feedback, water vapor feedback, and the decreasing trend of relative humidity with increasing resolution. Relatively small-scale downdrafts between upwelling columns over the substellar region are better resolved and the mixing between dry and wet air parcels and between anvil clouds and their environment are enhanced as the resolution is increased. These reduce atmospheric relative humidity and high-level cloud fraction, causing a lower clear-sky greenhouse effect, a weaker cloud longwave radiation effect, and subsequently a cooler climate with increasing model resolution. Overall, the sensitivity of the simulated climate of tidally locked aquaplanets to model resolution is small.

\end{abstract}



\section{Introduction} \label{introduction}

Tidally locked terrestrial planets around M dwarfs are the main targets of exoplanet missions for finding potentially habitable planets beyond Earth. This is due to their relatively large planet-to-star ratio and frequent transits. Several atmospheric general circulation models (AGCMs) have been modified to simulate the climate of potentially habitable, tidally locked planets \citep{merlis2010atmospheric,pierrehumbert2010palette,wordsworth2011gliese,edson2011atmospheric,yang2013stabilizing,menou2013water,turbet2016habitability,kumar2016inner,way2017resolving,wolf2017assessing,boutle2017exploring,paradise2017gcm,checlair2017no,noda2017circulation}. These studies generally agree that planets in the middle range of the habitable zone are able to maintain liquid water on the surface of the day side as long as when the atmosphere is not too thin. Further simulations using fully coupled atmosphere--ocean--sea-ice general circulation models (AOGCMs) showed that oceanic and sea-ice dynamics are able to modify the amount and spatial pattern of ice and snow and influence the climate \citep[e.g.,][]{hu2014role,del2019habitable,yang2020transition}.


Recent model intercomparison for tidally locked planets found that even under the same boundaries and the same stellar flux and atmospheric compositions, various AGCMs obtained significantly different results in planetary climate \citep{yang2019simulations,fauchez2020trappist}. The difference in surface temperatures can be as large as 20-30~K on the day side and 40-60 K on the night side among the models of CAM3, CAM4, ExoCAM, AM2, UM, ROCKE-3D, and LMD. Various aspects of the models can cause the divergence. The prime reasons are the differences in cloud parameterization and water vapor radiation transfer, as well as the interactions between these two processes and atmospheric dynamics \citep{yang2016differences, yang2019simulations}. In particular, differences in stellar absorption by water vapor can influence the height of last saturation points on the day side and subsequently affect atmospheric relative humidity \citep{yang2019simulations}. Differences in dry dynamical core have a very small effect on the results \citep{yang2019simulations}. Another possible reason that has not been examined is model resolution.

In previous simulations, different resolutions have been employed, such as 2.0$^{\circ}$$\times$2.0$^{\circ}$ with 32 levels in AM2 \citep{merlis2010atmospheric}, 3.8$^{\circ}$$\times$5.6$^{\circ}$ in latitude and longitude with 18 or 26 vertical levels in LMD \citep{wordsworth2011gliese,wordsworth2015atmospheric,turbet2016habitability}, 2.8$^{\circ}$$\times$2.8$^{\circ}$ or 3.75$^{\circ}$$\times$3.75$^{\circ}$ with 26 levels in the model CAM3 \citep{yang2013stabilizing, yang2014atmospheric, wang2016effects}, 1.9$^{\circ}$$\times$2.5$^{\circ}$ with 26 levels in CAM4 and CAM5 \citep{kumar2016inner,bin2018new}, 4$^{\circ}$$\times$5$^{\circ}$ with 40 levels in ExoCAM \citep{wolf2017assessing,wolf2017constraints} as well as in ROCKE-3D \citep{way2017resolving},  2$^{\circ}$$\times$2.5$^{\circ}$ with 40 levels in UM \citep{boutle2017exploring}, and 2.8$^{\circ}$$\times$2.8$^{\circ}$ or  5.6$^{\circ}$$\times$5.6$^{\circ}$ with 10 levels in PlaSim \citep{menou2013water,paradise2017gcm}. Whether the differences in model resolution has a strong or weak effect on the simulated results is unknown.

Atmospheric motions on Earth cover a wide range of space scales. In horizontal direction, the length scales are several to hundreds of meters for microscale turbulences, one kilometer to tens of kilometers for mesoscale convection systems such as thunderstorms and squall lines, hundreds of kilometers for synoptic phenomena such as tropical cyclones and fronts, and thousands of kilometers for macroscale phenomena such as baroclinic, standing, and tidal waves  \citep{orlanski1975rational}. Resolving all these scales is impossible and sub-grid parameterizations are necessary. Present atmospheric general circulation models for Earth generally have a grid spacing of $\mathcal{O}$(100) km, under which mid-latitude baroclinic and standing waves and large-scale circulations are resolved but convection, boundary layer mixing, and subgrid-scale waves and turbulences are required to be parameterized.

AGCM simulations of Earth climate have found that varying model resolution does not lead to drastic changes in mean climate such as global-mean surface temperature and precipitation, but it can influence the magnitudes and/or spatial patterns of precipitation, relative humidity, clouds, equatorial waves, jet streams, and air-sea interactions \citep{williamson2008convergence,williamson2013dependence,landu2014dependence,retsch2017vertical,retsch2019climate,vanniere2019multi}. For example, varying horizontal or vertical resolution can induce various locations for the the intertropical convergence zone (ITCZ, i.e., peak precipitation region in the tropics), from a single ITCZ at the equator to two ITCZs on either side of the equator. The underlying mechanisms are largely model dependent, which can be related to the better resolved mixing and vertical motion as the resolution increases or to resolution-dependent physics parameters. \cite{williamson2013dependence} showed that modifying vertical resolution in the AGCM CAM3 can influence the vertical distributions of relative humidity, specific humidity, cloud fraction, and longwave radiation cooling rate, which is due to the discrete approximation for the onset of shallow convection. \cite{landu2014dependence} addressed that horizontal resolution influences the distribution of specific humidity and subsequently the structure of ITCZ. \cite{retsch2017vertical} found that increasing vertical resolution leads to stronger mixing between updraft and its environment, which promotes an equatorward shift of the ITCZ. \cite{retsch2019climate} showed that relative humidity decreases with increasing horizontal resolution under both explicit and parameterized convection.

In this work, we explore the effect of model resolution on the simulated mean climate of tidally locked aquaplanets. The rest of this paper is organized as follows. Section 2 briefly describes the model and experimental designs, section 3 presents the results, and section 4 provides the summary. In the section 3, we will show the effect of varying horizontal resolution, the trend of relative humidity with increasing resolution (the key mechanism), and the effect of varying vertical resolution, respectively.



\section{Methods}\label{methods}

The atmospheric general circulation model used in this study is the Deep Paleo/Exoplanet CAM, which is called as ExoCAM (link: https://github.com/storyofthewolf). The model is based on the NCAR Community Atmospheric Model version 4 (CAM4) but modified by Eric Wolf for simulating early Earth and terrestrial exoplanets. Two main modifications are the radiation transfer for high concentrations of CO$_2$ and H$_2$O and the numerical solver for entropy calculation within the Zhang-MacFarlane convection parameterization \citep{wolf2013hospitable,wolf2015evolution,wolf2017constraints,wolf2017assessing}. Due to these two updates, ExoCAM is able to simulate moist and runaway greenhouse states, which is important for estimating the location of the inner edge of the habitable zone.

We performed seven experiments with five different resolutions under a finite-volume dynamical core: 1) 4$^{\circ}$$\times$5$^{\circ}$ in latitude and longitude for horizontal grids and 26 levels for vertical grids, labelled as f45\_L26; 2) 1.9$^{\circ}$$\times$2.5$^{\circ}$ and 26 levels, labelled as f19\_L26; 3) 0.5$^{\circ}$$\times$0.5$^{\circ}$ and 26 levels, labelled as f05\_L26; 4) 1.9$^{\circ}$$\times$2.5$^{\circ}$ and 51 levels, labelled as f19\_L51; and 5) 1.9$^{\circ}$$\times$2.5$^{\circ}$ and 74 levels, labelled as f19\_L74. The surface pressure is 1.0~bar in all the experiments. The maximum space between vertical levels is 96, 51, and 28 hPa for L26, L51, and L74, respectively. In the vertical direction, the scale height of water vapor in the atmosphere is about 2 km, and a vertical resolution of $\approx$25 hPa is required to well simulate the vertical distribution of water vapor \citep{tompkins2000vertical}. Among the experiments, all the parameterizations for sub-grid theromodynamics and dynamics and associated free parameters are the same, and the only difference is the resolution.

All the experiments are set to be 1:1 tidally locked, which means the rotation period is equal to the orbital period. By default, the stellar temperature is set to be 3,700 K, the stellar flux at the substellar point is 1,800 W\,m$^{-2}$, and  the orbital period is set to be 38.66 earth days; these parameters represent a typical slow rotation planet not far from the inner edge of the habitable zone around an early-M dwarf (ref.~Table~1 of \cite{kumar2017habitable}). Two experiments for a rapidly rotating aquaplanet are also performed with  resolutions of f45\_L26 and f05\_L26. The star temperature is 2,600 K, the stellar flux is 1,300 W\,m$^{-2}$, and the orbital period (= rotation period) is 4.25 Earth days; these parameters represent a typical rapidly rotating planet close to the inner edge of the habitable zone around a late-M dwarf. Stellar spectra are from the BT\_Settl stellar model \citep{allard2007k}.

The atmosphere is Earth-like, including N$_{2}$ and H$_{2}$O, but O$_2$, O$_3$, CO$_{2}$, CH$_{4}$ and aerosols are not considered in the simulations. Planetary radius and gravity are the same as Earth. The atmosphere is coupled to an immobile, slab ocean with a depth of 50 m everywhere; no continent is considered. Sea ice is allowed to form when the surface temperature is below the freezing point (271.35~K), and the albedos of sea ice and snow depend on the stellar spectrum. For the visible band ($<$0.7 $\mu$m), the snow albedo is 0.80 and the ice albedo is 0.67. For the near infrared band ($>$0.7 $\mu$m), it is 0.68 for snow and 0.30 for sea ice. Neither oceanic dynamics nor sea-ice dynamics are included in the simulations. By default, the time step of the experiments is 30 minutes, except for the f05\_L26, f19\_L51, and f19\_L74 experiments, within which the time step is 15 minutes to avoid numerical instability. Each experiment was run for tens of earth years until the sea ice and surface temperature reach equilibrium and the global-mean net energy imbalance at the top of the atmosphere (TOA) is within 1.0~W\,m$^{-2}$. Averages of the last 5 or 10 earth years were used here.


\section{Results}\label{results}

\subsection{Effect of Varying the Horizontal Resolution}\label{Hres}

\begin{figure*}
   \centering
   \includegraphics[width=1.0\linewidth]{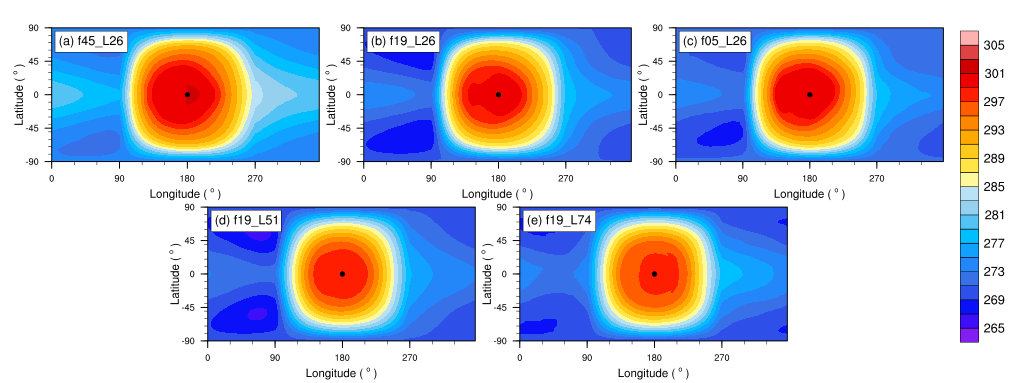}
   \caption{Surface temperature (K) in five experiments with resolutions of f45\_L26, f19\_L26, f05\_L26, f19\_L51, and f19\_L74 (see section~\ref{methods} for detailed descriptions). In all these experiments, the stellar flux is 1,800 W~m$^{-2}$, star temperature is 3,700~K, and rotation period (= orbital period) is 38.66 earth days.}
   \label{figure_TS}%
   \end{figure*}

\begin{figure*}
   \centering
   \includegraphics[width=1.0\linewidth]{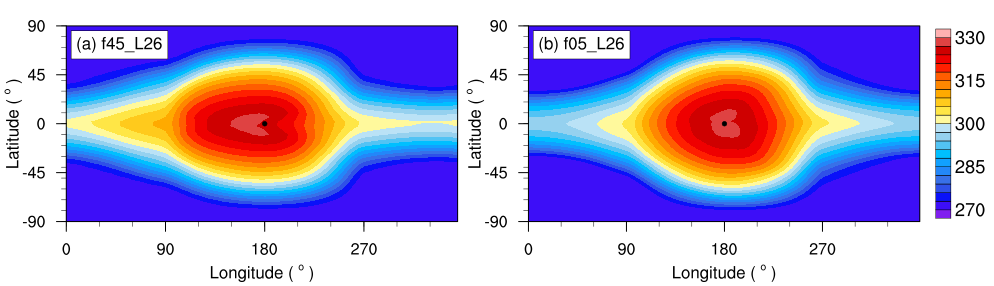}
   \caption{Surface temperature in the rapid rotation experiments with two resolutions of f45\_L26 (a) and f05\_L26 (b). In these two experiments, the stellar flux is 1,300 W~m$^{-2}$, the star temperature is 2,600~K, and the rotation period (= orbital period) is 4.25 earth days. The global-mean surface temperature is 294.7~K in (a) and 293.7~K in (b).}
   \label{figure_rapid}%
   \end{figure*}

Table~\ref{table1} summarizes the global-mean characteristics of the slow rotation experiments. Overall, varying the horizontal resolution has a small effect on the surface temperature (Fig.~\ref{figure_TS}(a-c)). In global mean, the surface temperature in the experiment of f45\_L26 is 3.7~K higher than that of f19\_L26 and 3.2~K higher than that of f05\_L26. In the rapid rotation experiments, the difference is also small (Fig.~\ref{figure_rapid}): the surface temperature difference between f45\_L26 and f05\_L26 is 1.0 K in global mean. In the following analyses, we focus on the results of slow rotation experiments, because the underlying mechanisms are the same.

\begin{table*}
\caption{Global-mean characteristics of the simulated climate of a slow rotation aquaplanet under five different resolutions in the model ExoCAM. For the definition of total relative humidity, please see text. The clear-sky greenhouse effect is defined as the difference in upward thermal infrared flux between the surface and the top of the model.}
\label{table1}   
\centering          
\begin{tabular}{l r r r r r}     
\hline\hline       
Variables & f45\_L26 & f19\_L26 & f05\_L26 & f19\_L51 & f19\_L74\\ 
\hline   
    surface temperature (K) & 284.5 & 280.8 & 281.3 & 279.6 & 280.3 \\  
    minimum surface temperature (K) &   271.1   &     267.2      &    268.0   &   266.6 & 268.7 \\
    maximum surface temperature (K) &   301.7   &    299.9        &    300.6   &   298.2 & 297.8 \\
    sea ice coverage &  4\% &  23\%  &  21\%    &  26\%  & 20\% \\
    surface albedo (0-1) & 0.065 & 0.066   & 0.062 & 0.064 & 0.064 \\
    planetary albedo (0-1) & 0.440 & 0.435   & 0.434 & 0.439 & 0.441 \\
    total relative humidity & 80.7\%   &  78.6\%  &  77.0\%  &   76.7\%   &   76.2\% \\
    clear-sky greenhouse effect (W/m$^{2}$) &  46.5 & 36.9  &  38.7  & 36.6  & 38.1   \\
    vertically-integrated water vapor (kg/m$^{2}$)  & 93.1  & 80.4   &  80.9  &  69.5  & 67.8 \\ 
    precipitation (mm/day) & 0.90 &  0.92  &   0.98    &   0.89 & 0.89 \\ 
    cloud shortwave radiation effect (W/m$^{2}$) &  $-$176.1 & $-$174.2  &   $-$173.7    &   $-$176.5 &  $-$177.6  \\
    cloud longwave radiation effect (W/m$^{2}$) &  47.1 &  37.4  &   39.1    &   36.6  & 38.1 \\
    vertically-integrated liquid cloud water (g/m$^{2}$) & 344.2 & 333.5   & 319.3 & 326.9 & 323.8 \\
    vertically-integrated ice cloud water (g/m$^{2}$) & 7.8 & 7.0 & 5.5 & 7.2 & 7.1  \\
    net energy imbalance at TOA (W/m$^{2}$)&   0.3   &   0.0   &  0.4     &   0.1 & -0.6   \\
\hline                  
\end{tabular}
\end{table*}

\begin{figure*} 
   \centering
   \includegraphics[width=1.0\linewidth]{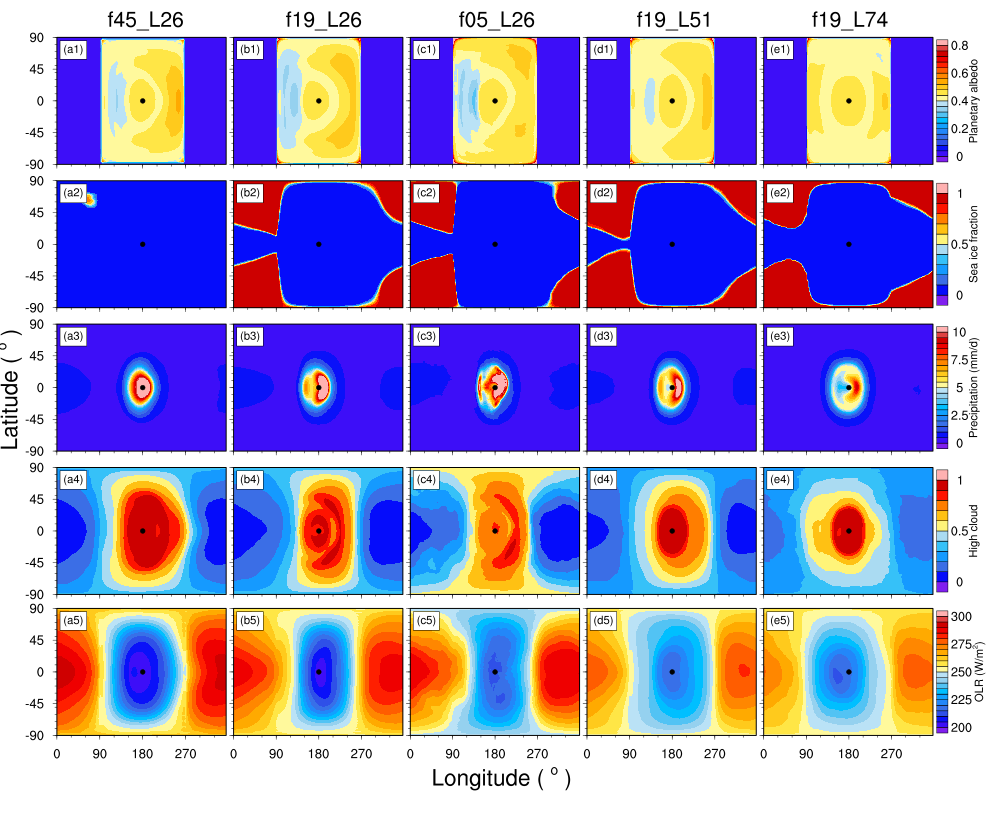}
   \caption{Planetary albedo (0-1, a1--e1), sea ice fraction (0-1, a2--e2), precipitation (mm/day, a3--e3), high-level cloud fraction (0-1, a4--e4), and outgoing longwave radiation at the top of the model (OLR, W\,m$^{-2}$, a5--e5) in the five slow rotation experiments with different resolutions: f45\_L26, f19\_L26, f05\_L26, f19\_L51, and f19\_L74. In each panel, the black dot is the substellar point.}
   \label{figure_albedo}%
   \end{figure*}

\begin{figure*}
   \centering
   \includegraphics[width=1.0\linewidth]{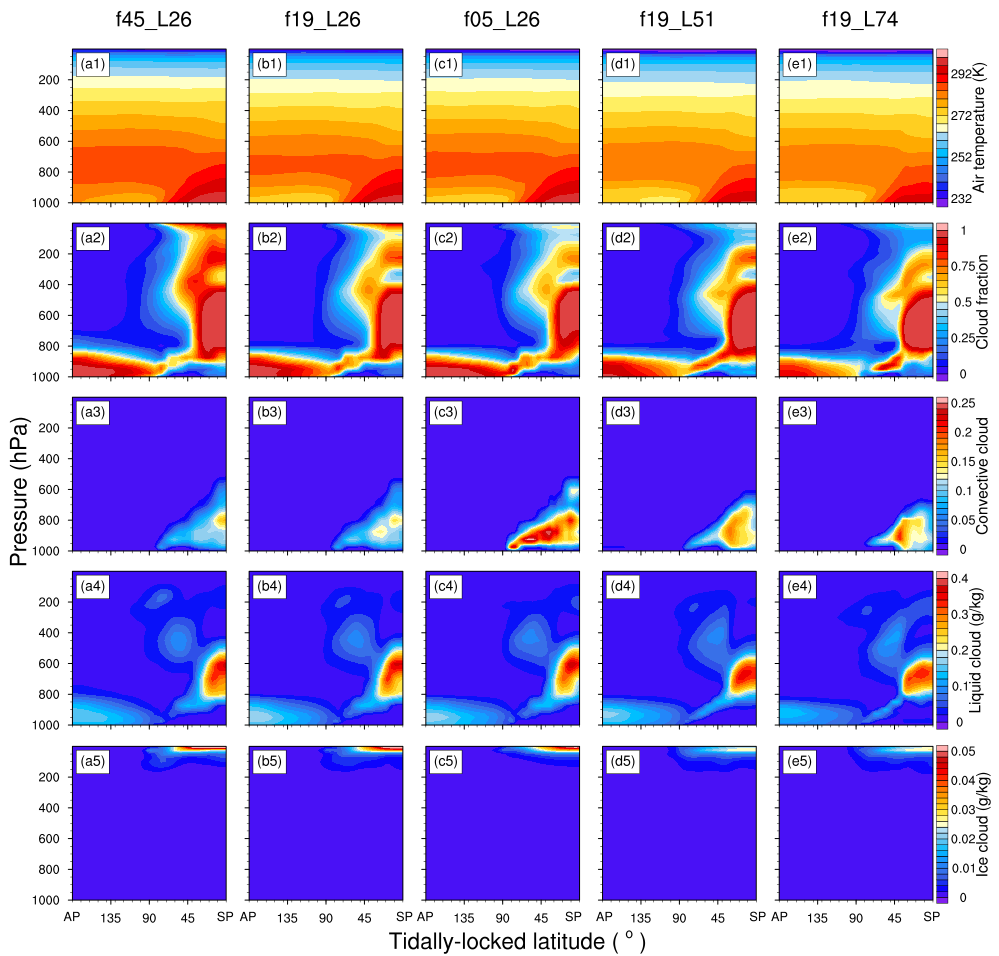}
   \caption{Air temperature (K, a1--e1), total cloud fraction (0-1, a2--e2), convective cloud fraction (0-1, a3--e3), liquid cloud amount (g/kg, a4--e4), and ice cloud amount (g/kg, a5--e5) in tidally locked coordinate in the five experiments with different resolutions: f45\_L26, f19\_L26, f05\_L26,  f19\_L51 and f19\_L74. AP: antistellar point, and SP: substellar point. For the tidally locked coordinate, please see \cite{koll2015deciphering} and https://github.com/ddbkoll.}
   \label{figure_T}%
   \end{figure*}

For planetary albedo, the difference is also small (Fig.~\ref{figure_albedo}(a1-c1)). The f45\_L26 experiment has a planetary albedo of 0.440, being 0.005 higher than that of f19\_L26 and 0.006 higher than that of f05\_L26, which corresponds to a difference in global-mean shortwave reflection of 2.25 and 2.70 W\,m$^{-2}$, respectively. The cloud longwave radiation effects are 47.1, 37.4, and 39.1 W\,m$^{-2}$ in the experiments of f45\_L26, f19\_L26, and f05\_L26, respectively. \textcolor{black}{The smaller planetary albedo and lower cloud longwave radiation effect as increasing the resolution are from lower relative humidity (discussed below), less high-altitude cloud coverage (Figs.~\ref{figure_albedo}(a4-c4) \&~\ref{figure_T}(a2-c2)) and smaller cloud water path (Fig.~\ref{figure_T}(a4-c4) \&~(a5-c5)) especially above the level of 400~hPa over the substellar region.} In global mean, the vertically-integrated liquid cloud water paths are 344.2, 333.5, and 319.3 g\,m$^{-2}$ and vertically-integrated ice cloud water paths are 7.8, 7.0, and 5.5 g\,m$^{-2}$ in the experiments of f45\_L26, f19\_L26, and f05\_L26, respectively (Table~\ref{table1}). Moreover, cloud height does not exhibit significant trend in these experiments (figure not shown). Another clear trend is that convective cloud fraction increases with resolution (Fig.~\ref{figure_T}(a3-c3)), due to the fact that a higher model resolution enables the generation of more explicit convective plumes.

The largest difference is in sea ice fraction, as shown in Fig.~\ref{figure_albedo}(a2-c2). In the experiment of f45\_L26 there is nearly no ice on the night side, whereas the night side in the experiments of f19\_L26 and f05\_L26 is covered by sea ice except in the low latitudes. The global-mean surface coverage is 4\%, 23\%, and 21\%, respectively. The difference is due to the fact that the night-side surface temperatures in f45\_L26 are several degrees higher than those in the other two experiments (see Fig.~\ref{figure_TS}(a-c)). Sensitivity tests using different initial temperatures show that these results are independent on the initial state (figure not shown). Note that the night-side surface temperatures are roughly uniform and are close to the freezing point (271.35~K in the model) in all these three experiments under the stellar flux of 1,800 W\,m$^{-2}$, so that a small warming of the surface is able to melt a large area of ice. If the stellar flux were set to be higher or lower than 1,800 W\,m$^{-2}$, difference in sea ice fraction among the experiments would be smaller.

Note that the change of sea ice coverage on the night side does not exhibit an ice-albedo feedback on the surface temperature because there is no stellar radiation on the permanent night side.  Moreover, there is no significant difference in surface albeodo between these experiments. The sea-ice and snow albedos are spectrally dependent as addressed in the above section, so that the surface broadband albedo depends on the partitioning between near-infrared and visible solar radiation at the surface. This, in turn, depends on cloud cover and atmospheric humidity associated with shortwave scattering and absorption, and is therefore climate-state dependent. This is the reason why the relatively warmer  f45\_L26 experiment does not have a smaller surface albedo than other cases.

The sea ice fraction in the case of f19\_L26 is much higher than that of f45\_L26 but quite similar to that of f05\_L26, indicating a sign of convergence. This result suggests that in order to well simulate the sea ice coverage, a minimum horizontal resolution of $\approx$200~km is necessary at least in the model employed here.

The bottom row of Fig.~\ref{figure_albedo} shows that the outgoing longwave radiation at the top of the model (i.e., thermal infrared emission to space) on the day side is lower than that on the night side, although the surface temperatures on the day side are much greater than on the night side. This is mainly due to the effect of day-side clouds especially the high-level clouds (Fig.~\ref{figure_albedo}(a4-c4)) that absorbs the high thermal radiation from the surface but emits thermal energy to space at temperatures of the cloud top, which are much lower than that of the surface. This thermal reverse between day and night sides were first found in the simulations of \cite{yang2013stabilizing} and then confirmed in many other studies such as \cite{kumar2016inner}, \cite{haqq2018demarcating}, and \cite{wolf2019simulated}; future space telescopes such as the \textit{James Webb Space Telescope} may be able to observe this thermal emission reverse especially on planets close to the inner edge of the habitable zone. Note that the outgoing longwave radiation on the day side becomes somewhat larger as increasing the horizontal resolution; this is because the high-level cloud coverage decreases with resolution as discussed above.

Analyses of zonal-mean zonal winds show that atmospheric superrotation (i.e., west-to-east flows in the tropical region) has no large change as varying the horizontal resolution (figure not shown). This is consistent with the small sensitivity of surface temperature. Note that the asymmetries in sea ice fraction, cloud distribution, and precipitation between the west and the east of the substellar point (see Fig.~\ref{figure_albedo}) are due to the atmospheric superrotation.

\subsection{The Decreasing Trend of Relative Humidity with Resolution}

For the differences addressed in section~\ref{Hres}, the main reasons are the effect of cloud feedback (addressed above) and the decrease of relative humidity with resolution (addressed in this section). Water vapor feedback acts to further amplify the differences. The decreasing trend of relative humidity with resolution is due to better-resolved small-scale downdrafts in the substellar mean upwelling region and enhanced mixing between dry and wet parcels, which are schematically summarized in Fig.~\ref{fig_mechanism}(a-b).

\begin{figure*} 
   \centering
   \includegraphics[width=1.0\linewidth]{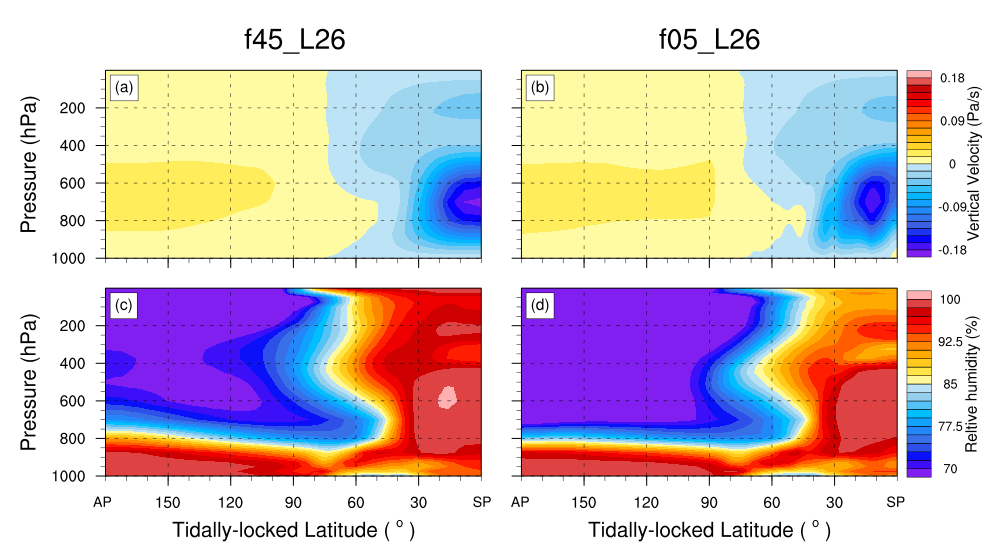}
   \caption{Long-term mean vertical velocity (negative value represents upwelling and positive downwelling, a-b) and relative humidity (c-d) in tidally locked coordinate in the slow rotation experiments of f45\_L26 (left) and f05\_L26 (right). As the resolution is increased, the mean upwelling region becomes narrower in width and relative humidity decreases. AP: antistellar point, and SP: substellar point.}
   \label{figure_RH}%
   \end{figure*}

\begin{figure*} 
   \centering
   \includegraphics[width=1.0\linewidth]{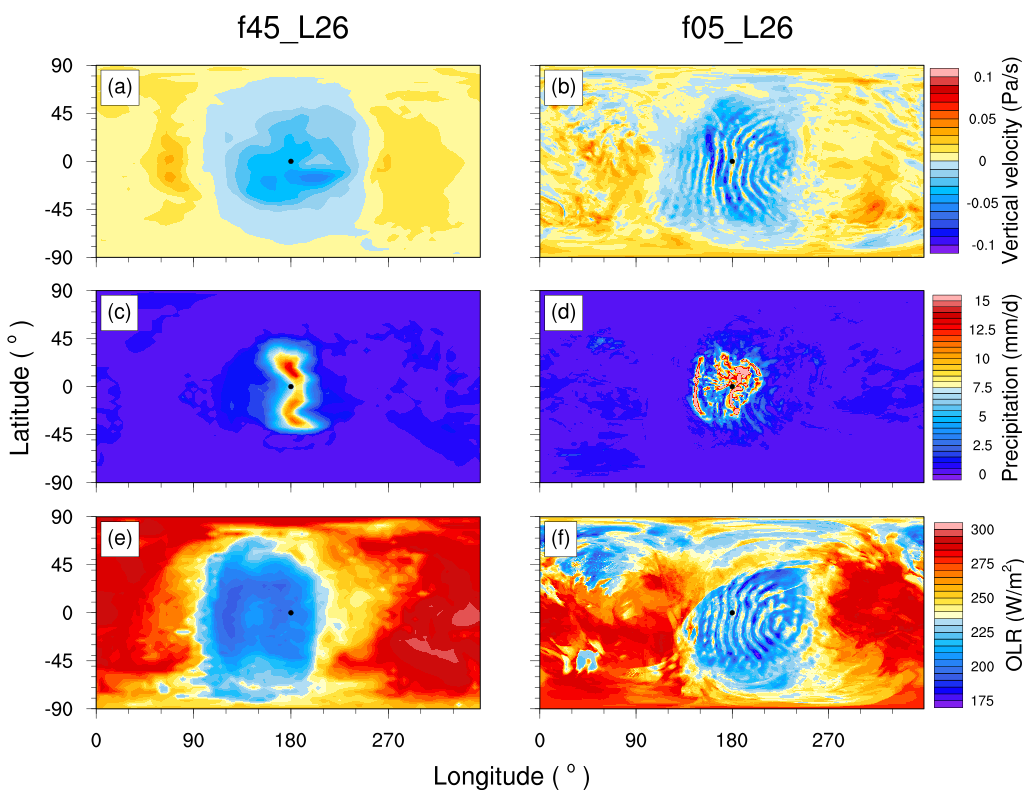}
  \caption{Snapshots of vertical velocity at 300 hPa (a-b), precipitation (c-d), and outgoing longwave radiation at the top of the model (e-f) in the slow rotation experiments of f45\_L26 (left) and f05\_L26 (right). As the resolution is increased, smaller-scale updrafts, downdrafts, and waves can be better resolved.}
  \label{figure_snapshot}%
  \end{figure*}

\begin{figure*} 
   \centering
   \includegraphics[width=1.0\linewidth]{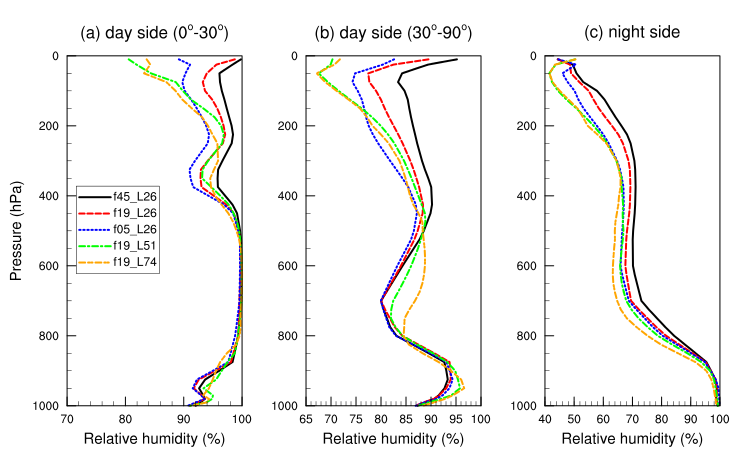}
   \caption{Area-mean relative humidity profiles over the substellar region between 0$^{\circ}$--30$^{\circ}$ (a), between 30$^{\circ}$--90$^{\circ}$ on the day side (b), and on the entire night side (c) in tidally locked coordinate of the slow rotation experiments.}
   \label{figure_RHprofile}%
   \end{figure*}

Atmospheric circulation on the slowly rotating, tidally locked planet is dominated by a global-scale Walker circulation with mean upwelling over the substellar region and mean downwelling in the rest of the planet (Fig.~\ref{figure_RH}(a)). Over the substellar region, the relative humidity is high due to convection and large-scale convergences. At the high latitudes of the day side and on the entire night side the relative humidity is low due to large-scale downwelling, except near the surface for which surface evaporation constantly moistens the atmosphere in the planetary boundary layer (Fig.~\ref{figure_RH}(c)). As the horizontal resolution is increased, relatively small-scale waves and downwelling columns in the background upwelling region on the day side can be better resolved (Fig.~\ref{figure_snapshot}). Because the downwelling columns are relatively dry, the mean relative humidity decreases (Fig.~\ref{figure_RH}(d)). On the night side, the relative humidity also decreases (Fig.~\ref{figure_RHprofile}(c)), due to that less water vapor can be supplied to transport from the day side to the night side, although the horizontal winds are effectively unchanged (figure not shown). Around the terminators between the mean upwelling and mean  downwelling regions, relative humidity exhibits the same decreasing trend (Figs.~\ref{figure_RH}(d) \&~\ref{figure_RHprofile}(b)). This is due to the enhanced horizontal mixing between dry and wet parcels as the model resolution is increased. Moreover, the mixing between air parcels and high-level anvil clouds over the substellar region becomes stronger as the resolution is increased (schematically shown in Fig.~\ref{fig_mechanism}(c-d)), so that the anvil clouds are more difficult to be maintained and the high-level cloud fraction decreases as addressed in section~\ref{Hres}.

A clear trend in the long-term mean vertical velocity is the shrinking of the strong upwelling region over the substellar region as shown in Fig.~\ref{figure_RH}(b)\,vs.\,(a), or in other words the strong upwelling region becomes narrower in width. This is due to the combination of the resolved downdrafts in the mean upwelling flows over the substellar region and the enhanced lateral mixing of dry and wet parcels between the mean downwelling and mean upwelling regions.

As shown in Fig.~\ref{figure_RHprofile}, the decreasing of relative humidity on the day side mainly occurs between the level of $\approx$400~hPa and the top of the model, and on the night side it occurs between the level of $\approx$800~hPa and the top of the model. Below the level of $\approx$400~hPa on the day side,  relative humidity nearly does not change because this region has robust convective plumes and large-scale convergences and the atmosphere is close to saturation in all the cases. Below the level of $\approx$800~hPa on the entire night side, the relative humidity is also close to saturation in all the cases because water vapor evaporated from the surface is trapped under the temperature inversion (see Fig.~\ref{figure_T}(a1-c1)).

\begin{figure*} 
   \centering
   \includegraphics[width=1.0\linewidth]{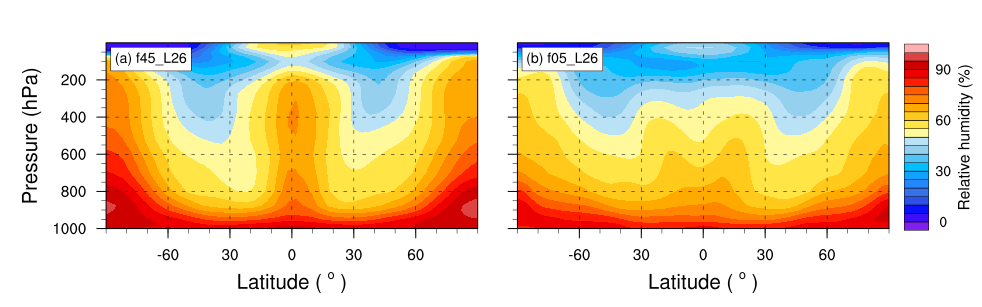}
   \caption{Zonal-mean relative humidity (\%) in the rapid rotation experiments with resolutions of f45\_L26 (a) and f05\_L26 (b). The relative humidity tends to decrease as the resolution increases.  The total relative humidity is 61.4\% and 58.4\%, respectively.}
   \label{figure_RH3}%
   \end{figure*}

The global-mean relative humidity at the level of 200~hPa is 76.9\%, 73.3\%, and 68.4\% and total relative humidity is 80.7\%, 78.6\%, and 77.0\% in the experiments of f45\_L26, f19\_L26, and f05\_L26, respectively. Following \cite{wolf2015evolution}, the total relative humidity is defined as the percentage of water vapor by mass contained in the whole atmosphere compared with the water vapor mass that the atmosphere could theoretically hold if saturated everywhere. In the two rapid rotation experiments of f45\_L26 and f05\_L26, the mean relative humidity also decreases (Fig.~\ref{figure_RH3}). The decreasing trend of mean relative humidity with model resolution has also been mentioned in Earth simulations of \cite{retsch2019climate} and \cite{vanniere2019multi}. The decreased relative humidity reduces specific humidity and diminishes atmospheric greenhouse effect under given air temperatures; water vapor feedback acts to further amplify the difference in surface temperature. In these three experiments, the vertically integrated water vapor amounts are 93.1, 80.4, and 80.9 kg\,m$^{-2}$ and the clear-sky greenhouse effects are 46.5, 36.9, and 38.7~W\,m$^{-2}$, respectively (Table~\ref{table1}).

\begin{figure*} 
   \centering
   \includegraphics[width=1.0\linewidth]{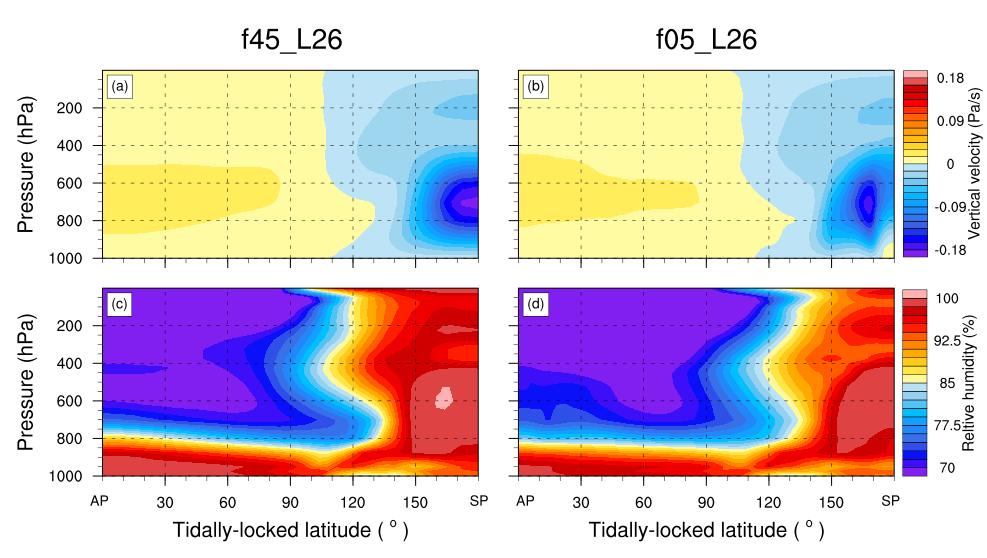}
   \caption{Long-term mean vertical velocity (a--b) and relative humidity (c--d), same as Fig.~\ref{figure_RH} but for the fixed surface temperature experiments under resolutions of f45\_L26 (left) and f05\_L26 (right). As the resolution is increased, the mean strong upwelling region shrinks and the mean relative humidity decreases. In these two experiments, surface temperatures are obtained from the slab-ocean experiment shown in Fig.~\ref{figure_TS}(a).}
   \label{figure_RH2}%
   \end{figure*}

In order to further confirm the mechanism for the effect of varying resolution on relative humidity, we added two experiments within which surface temperatures are fixed. Because differences in surface temperatures can also influence the degree of atmospheric sub-saturation and its spatial pattern \citep[e.g.,][]{lau2015robust}, these two experiments are able to eliminate the effect of the divergence in surface temperatures. Two resolutions were run, f45\_L26 and f05\_L26, and the results are shown in Fig.~\ref{figure_RH2}. Again, the relative humidity decreases with resolution, especially in the terminator regions between the mean upwelling and mean downwelling regions and in the substellar region above the level of $\approx$400 hPa. The global-mean relative humidity at the level of 200~hPa is respectively 76.3\% and 70.0\% and the total relative humidity is 80.5\% and 78.0\% in these two experiments. Note that the magnitude of the decreasing of relative humidity in the fixed surface temperature experiments is slightly smaller than that in the coupled slab-ocean experiments. The underlying reason is likely due to the lack of the coupling between the surface and atmospheric dynamics in these two experiments.

\subsection{Effect of Varying the Vertical Resolution}

Varying the vertical resolution also has a small effect on the climate. The global-mean surface temperatures are 280.8, 279.6, and 280.3~K and the global-mean planetary albedos are 0.435, 0.439, and 0.441 in the cases of f19\_L26, f19\_L51, and f19\_L74, respectively (Table~\ref{table1}). The global-mean sea ice coverage is also similar, 23\%, 26\%, and 20\%.

\textcolor{black}{The effect of increasing vertical resolution on the relative humidity is similar to the effect of increasing horizontal resolution, as shown in Fig.~\ref{figure_RHprofile}. The relative humidity decreases significantly in the upper troposphere over the substellar region, due to the better resolved downdrafts in the background upwelling region and to the enhancement of mixing between dry and wet parcels. The fraction of high-level clouds also decreases (Figs.~\ref{figure_albedo}(b4, d4, \& e4) \& \ref{figure_T}(b2, d2, \& e2)), due to the combination of reduced relative humidity and increased mixing between anvil clouds and their environment. This mechanism is similar to that in the Earth simulations of \cite{retsch2017vertical}. The global-mean relative humidity at 200~hPa is 73.3\%, 69.0\%, and 66.4\% and the total relative humidity is 78.6\%, 76.7\%, and 76.2\% in the experiments of f19\_L26, f19\_L51, and f19\_L74, respectively. As a result of the decreasing trends of relative humidity and anvil clouds, the outgoing longwave radiation to space raises slightly (Fig.~\ref{figure_albedo}(b5, d5, \& e5)).}



\section{Summary}\label{summary}

\begin{figure*} 
   \centering
   \includegraphics[width=0.8\linewidth]{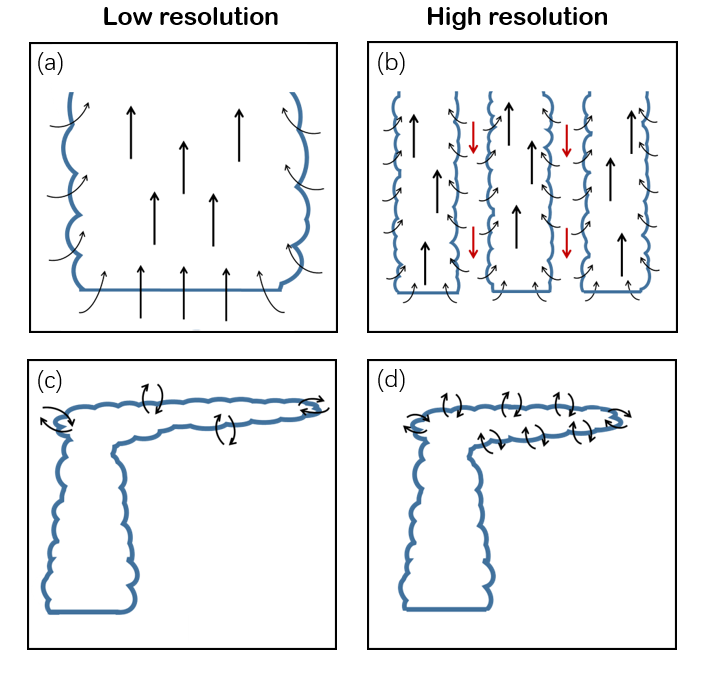}
   \caption{Schematic diagram for the effects of increasing model resolution. (a) versus (b): Subsidence between convective plumes is better resolved and the mixing between dry and wet parcels increases. (c) versus (d): The mixing between anvil clouds and their environment enhances. These changes reduce the mean relative humidity and high-altitude cloud fraction, causing a smaller clear-sky greenhouse effect and a lower cloud longwave radiation effect. Therefore, the surface temperature decreases with increasing resolution.}
   \label{fig_mechanism}%
   \end{figure*}

The 3D AGCM ExoCAM is used to employ the influence of model's horizontal and vertical resolutions on the simulated climate of tidally locked aqua-planets around low-mass stars. Three main conclusions are as follows: 

   \begin{enumerate}
    \renewcommand{\labelenumi}{(\theenumi)}
      \item The mean climate depends on model resolution slightly. In global mean, the difference is within 5~K in surface temperature, 0.007 in planetary albedo, and 3.2~W\,m$^{-2}$ in net energy flux. These values are much smaller than the differences of 20$-$30~K in global-mean surface temperature, 0.16 in planetary albedo, and 10$-$20 W\,m$^{-2}$ in water vapor radiative transfer   shown in \cite{yang2016differences,yang2019simulations} and \cite{fauchez2020trappist}.
      \item The global-mean surface temperature decreases with increasing resolution. As increasing the model resolution, small-scale downdrafts between upwelling columns can be better resolved and the mixing between dry and wet parcels and between air parcels and anvil clouds become stronger (Fig.~\ref{fig_mechanism}). These act to reduce the mean relative humidity and decrease the high-level cloud fraction. As a result, both clear-sky greenhouse effect and cloud longwave radiation effect decrease, leading to a cooler surface. But, convective cloud fraction increases, due to the fact that more convection cells are resolved as the resolution is increased.
      \item As varying the resolution, the concentration of surface sea ice starts to converge at a resolution of 210~km. This result suggests a minimum resolution of $\approx$200~km is necessary in the model ExoCAM, in order to better simulate the climate. 
   \end{enumerate}

No continent is involved in our experiments. Earth simulations showed that a higher resolution is required to well resolve clouds and precipitation over topography (such as \citealt{lau2009simulation}). Moreover, as increasing model resolution, rainfall decreases over ocean but increases over land due to the increases in atmospheric moisture transport from ocean to land (such as \citealt{demory2014role}). This implies that for tidally locked planets with continents a higher resolution is required.


\acknowledgments

We are grateful to Eric Wolf for the release of the model ExoCAM, to Shaozhi Lin for his help in model setup, and to Feng Ding and Wanying Kang for helpful discussions. J.Y. acknowledges support from the National Natural Science Foundation of China (NSFC) grants 41761144072 and 41861124002.



\bibliography{mybib}{}
\bibliographystyle{aasjournal}



\end{document}